# Surface Adhesion of Back-Illuminated Ultrafast Laser-Treated Polymers

Deepak L. N. Kallepalli*, Alan T. K. Godfrey, Jesse Ratté, André Staudte, Chunmei Zhang, and P. B. Corkum*

Joint Attosecond Science Laboratory, University of Ottawa, and the National Research Council of Canada, 100 Sussex Dr., Ottawa K1N 6N5, Canada

Deepak.Kallepalli@uottawa.ca, pcorkum@uottawa.ca

## Abstract

We report a decreased surface wettability when polymer films on a glass substrate are treated by ultra-fast laser pulses in a back-illumination geometry. We propose that back-illumination through the substrate confines chemical changes beneath the surface of polymer films, leaving the surface blistered but chemically intact. To confirm this hypothesis, we measure the phase contrast of the polymer when observed with a focused ion beam. We observe a void at the polymer-quartz interface that results from the expansion of an ultrafast laser-induced plasma. A modified polymer layer surrounds the void, but otherwise the film seems unmodified. We also use X-ray photoelectron spectroscopy to confirm that there is no chemical change to the surface. When patterned with partially overlapping blisters, our polymer surface shows increased hydrophobicity. The increased hydrophobicity of back-illuminated surfaces can only result from the morphological change. This contrasts with the combined chemical and morphological changes of the polymer surface caused by a front-illumination geometry.

## 1. Introduction

The use of ultrafast pulses for local energy deposition in materials has several applications in 3D optical data storage [1-2], integrated optics [3-5], and Laser-Induced Forward Transfer (LIFT) [6-11]. In contaminant-free LIFT, the nonlinear interaction of an ultrafast pulse with the polymer film (often called a dynamic release layer (DRL)) between a glass substrate and transfer material, ensures confinement of chemical changes to the glass-polymer interface, leaving the transfer material unaffected [12]. When an ultrafast pulse (duration of ~ a few femtoseconds) interacts with a polymer film at the interface, it creates a localized hot plasma [13-14]. The hot plasma expands leading to formation of a blister.

The adhesive properties of materials are altered by both front- and back-illumination. However, front-illumination induces both morphological and chemical changes to the surface [15-16]. Front-illuminated surfaces can lead to hydrophilic, hydrophobic, and super-hydrophobic states [17-25]. Thus, changes to surface adhesion are often attributed to an interplay between surface chemistry and morphology. To the best of our knowledge there have been no reports of changes to adhesion induced by an ultrafast laser in a back-illuminated geometry.

We report an increase in hydrophobicity of a polyimide film induced by an ultrafast laser in a back-illumination geometry. Using focused ion beam (FIB) microscope and X-ray photoelectron spectroscopy (XPS), we find that these changes are strictly morphological at the interface of the film. This observation is consistent with our previous spectroscopic measurements, which showed that laser-induced chemical changes are confined to the glass-polyimide interface [12]. We confirm the increased hydrophobicity by measuring the contact angle of water droplets on laser-modified surfaces and we find that the contact angle increases with laser fluence. Thus, LIFT can be controlled by the energy content of a short pulse and/or by controlling the adhesion properties of a surface on which the object to be LIFTed resides.

2. **Experiment**

We prepared polyimide films on two different substrates: (1) on #1.5 Fisherbrand borosilicate glass coverslips (for FIB and contact angle measurements) and (2) on 500 μm fused silica discs for our XPS measurements. Both were rinsed with acetone, isopropanol, and deionized water to remove contaminants and dried on a hotplate. Polyimide films were made using PI-2525 (for thicker films) and PI-2555 (for thinner films < 1.4 μm) precursors from HD Microsystems following the recommended spin curves and baking conditions.

We used a Ti:Sapphire laser (Coherent RegA 9040) producing pulses of 50-fs duration at a central wavelength of 800 nm. The laser pulse energy was controlled using a motorized half-wave plate followed by a polarizing beam-splitter cube. The laser beam was focused using microscope objectives (10x 0.2 NA, 20x 0.4 NA) mounted into a vertical motor stage (PI M-112) with a travel range of 25 mm for adjusting the focal spot on the sample. Since polyimide is transparent at a wavelength of 800 nm (1.55 eV), any modification is due to nonlinear absorption.

We mounted polyimide-on-glass coverslips onto a 5-axis piezo nano-precision stage (PI) assembled on top of a micro-precision horizontal XY stage (MICOS MS-4). The nano-precision stage was used for fine adjustment of the focal position. The laser was focused through the glass substrate onto the glass-polymer interface (back-illumination geometry). A dichroic mirror was used before the microscope objective in a coaxial geometry, allowing a small portion of the focused laser light to back-reflect from the sample, re-collimate through the objective, and travel to an imaging line for in-situ laser spot monitoring. Coupling in white light and changing the position of the objective also allows for in-situ white-light microscopy. We used this to find the optimal position of the laser focus on the sample, by firing pulses with energies near the damage threshold while adjusting the position of the focal spot. Characterization details are in **Supplemental section I.**

**3. Results and discussion**

The self-focusing threshold in the substrate is P = 2.8 MW and for a 50-fs pulse, this corresponds to 140 nJ energy. Despite the losses in the substrate, the remaining energy delivered to the polymer film at the interface leads to a blister formation. Increasing the energy, especially for a low NA lens, will lead to continuum generation and breakdown in the substrate, but still the blister expands [13]. In this section, we report pulse energies delivered through the substrate accounting for the nonlinear absorption in the substrate.

We carried out laser irradiation experiments using polyimide thin films coated on glass and examined the changes in chemical composition and morphology at the glass-polymer interface and the surface. Contaminant-free LIFT requires confinement of chemical changes near the glass-polyimide interface while the polymer surface undergoes a morphological change through blister formation. Since polymers possess lower surface energies (~ 40 mJ/m$^2$ for polyimide) compared to metals (~ J/m$^2$), we chose a test polymer and carried out studies on chemical composition and morphology [26-28]. We used FIB and XPS measurements to study the morphology and chemical composition. In later sections, we show hydrophobicity test results on laser-irradiated polyimide surfaces to provide its relevance to the LIFT technique and other applications.

### 3.1 CHEMICAL COMPOSITION OF THE SURFACE

In this section, we present results from FIB measurements that image changes near the interface, followed by XPS measurements. For FIB analysis, we fabricated a series of individual blisters with ~ 130 nJ of pulse energy focused by a 0.4 NA microscope objective reaching an intensity of $5.3 \times 10^{13}$ W/cm$^2$ at the interface. All laser-irradiation experiments were carried out by focusing pulses at the glass-polyimide interface.

We used focused gallium (Ga) and helium (He) ion beams to dissect and image the interior of blisters, respectively as shown in **Fig. 1(a)**. The dissected blister showed a thin embedded layer of modification underneath the unmodified polymer in the blister. The image contrast indicates that the layer must have undergone chemical transformation due to nonlinear absorption of 800 nm light. In addition, a void is present beneath the embedded layer. It is partially filled with molten material that has solidified.

This observation confirms what we expect. When an ultrafast laser pulse is focused on the material interface, it forms a plasma [12-13]. At the temperature that we achieve, the material vaporizes and undergoes a chemical change creating the embedded layer and the solidified material on glass, as seen in **Fig. 1(a)**. Earlier reports involving blister formation in thin films of polyimide and titanium using 355 nm and 800 nm laser wavelengths, lack the direct experimental evidence of confined chemical changes at the interface [5, 11, 29, 30]. Our observation of confined chemical changes provides direct experimental evidence for confinement of chemical changes only at the interface induced by an ultrafast laser [5-8, 29-41].

XPS is a surface characterization technique with a typical sampling depth of 7.5 nm [42, 43]. Hence, we chose XPS to study any chemical changes that occurred to the front surface in back-illuminated experiments on polyimide. Since XPS characterization requires a large, patterned surface area, we patterned 2 mm × 3 mm areas on 1.3 μm thin films of PI-2555 by overlapping the individual blisters in back-illumination geometry. Two regions were modified with 870 nJ (high irradiation dose) and 330 nJ (low irradiation dose) pulse energies focused by a 0.2 NA objective at a fixed laser repetition rate of 500 Hz. The scan speeds and line spacings for these regions were 4.5 mm/s and 3 mm/s and 12 μm and 8 μm, respectively. The films experienced delamination and breakage at energies exceeding 500 nJ with a line spacing of less than 12 μm, which means water droplets could not be suspended on the surface for hydrophobicity tests.

**Figure 1(b)** shows the C (1S) envelope for pristine and laser-irradiated samples (black, red and blue for high and low irradiation dosed polyimide). For pristine polyimide, the carbon envelope

(black curve in **Fig. 1(b)**) consists of two peaks at 288.35 eV and 284.85 eV corresponding to C=O and C-C/C-H respectively. The XPS spectra recorded for laser-modified polyimide surfaces (red and blue curves in **Fig. 1(b)**) were corrected for energy shifts due to charge compensation (~ 2 eV) and plotted to compare with pristine polyimide [23]. There were no peak shifts observed, indicating that the surface is chemically intact after the laser treatment. We also compared the intensity ratios of peaks at 284.85 eV (C-C/C-H) and 288.35 eV (C=O) for laser-modified polyimide samples (blue and red curves) with the pristine polyimide. For pristine polyimide, the ratio of the intensities of C-C (24,143 cps) and C=O (5,900 cps) was around 4. For laser-modified polyimide with high irradiation dose, the ratio of intensities of C-C (36,458 cps) and C=O (9,660 cps) was 3.8 and for the low-irradiation dose polyimide sample, the ratio of intensities of C-C (29,208 cps) and C=O (7,224 cps) was 4. From this, we conclude that the carbon bonds at the surface of film were not altered. We also compared O(1S) and N(1S) envelopes for laser-modified polyimide with pristine and did not observe a peak shifts or significant changes in intensity ratios (Details are in **Supplemental section II**), further supporting that the surface chemistry was not altered.

### 3.3 ROLE OF SURFACE MORPHOLOGY ON ADHESION

Having determined that back-illuminated ultrafast- laser-treated-polymers leave the chemical composition of the polymer intact, we next turn to the role of surface morphology (roughness) on surface adhesion. It is known that surface adhesion is influenced by both chemistry and morphology. When a water droplet is placed on a flat surface (zero roughness), it shows a contact angle known as Young's angle ($\Theta_0$) illustrated in **Fig. 2(a)**. The contact angle results from a balanced surface force between three interfaces: air-solid, solid-liquid, and liquid-air.

The initial chemical state of a polymer surface depends on the baking conditions (such as temperature or time). When polymers are baked for a long time or at higher temperatures, hydrophilic groups are removed resulting in enhanced hydrophobicity. **Figure 2(b)** and **Figure 2(c)** show the contact angles of 110° and 82° for water droplets placed on polyimide thin films prepared at baking temperatures of 300°C and 180°C for 30 mins, respectively.

We irradiated samples prepared at a baking temperature of 300°C for 30 mins. When roughness is added to any substrate, it increases its initial hydrophilic (Wenzel) or hydrophobic (Cassie-Baxter) state. Since our experiments are carried out in back-illumination geometry on polymers prepared at 300°C, the addition of surface roughness enhances hydrophobicity as shown in **Fig. 2(d)**. In these experiments, the surface roughness induced by an ultrafast laser is nonuniform due to the intensity distribution of the pulse, in contrast to the uniform roughness, shown in **Fig. 2(d)**.

Since a 1 μL water droplet requires a ~ mm$^2$ surface area for a hydrophobicity measurement, we patterned surface areas of 2 mm × 3 mm under a $10^{-3}$ Torr vacuum (to avoid interaction with atmospheric oxygen [24]). The estimated spot size ($\omega_0$) (and the corresponding size of modification) is given by $\omega_0 = 1.22 \times \lambda/NA \times n^{0.5}$ where $\lambda$ is the wavelength of the laser (800 nm), NA is the numerical aperture of the microscope objective (0.4), and n is the minimum number of photons required for nonlinear absorption (2 for polyimide) [44]. From this equation, the estimated spot size is 1.8 μm for a 0.4 NA objective (neglecting the laser pulse energy (E) and thermal properties of the material). For our experiments, we varied the speed of the translation stage and the pulse energy at laser repetition rates of 500 Hz and 2 kHz to fabricate equivalent single-shot and overlapped blister patterns. The number of equivalent laser shots (N) per focal spot of diameter ($\omega_0$) and at scan speed (v) is given by the equation N = $\omega_0 \times$ f/v [45].

We patterned surfaces with 1, 2, and 4 shots per focal spot diameter on average, by varying scan speed (v), laser repetition rate (f), and line spacing ($\Delta X$ shown in **Fig. 3**). AFM topographies of the patterned surfaces are shown in **Fig. 3**. Pristine polyimide had an average surface roughness (R) of 0.2 nm and a surface height of 8 nm. Patterned surfaces with N = 1, 2, and 4 had average roughnesses of 225 nm, 125 nm, and 59 nm with surface heights of 1.2 µm, 800 nm, and 350 nm, respectively. The maximum patterned surface heights for N = 4 and 2 were less than the maximum heights for N = 1 because the energy (E) used to pattern these surfaces was varied with the number of shots (N) to avoid rupture of the film.

The addition of laser-induced surface roughness to an existing hydrophobic surface (**Fig. 4(a)**) transforms it into a super-hydrophobic surface by minimizing the liquid contact area fraction (f) given by cos($\Theta_{CB}$) = f(1 + cos($\Theta_0$))-1 [46,47]. The creation of laser-induced surface roughness creates more air pockets which further enhance hydrophobicity. From the initial state of pristine polyimide with a contact angle of 110°, we plotted the liquid contact area fraction (f) as shown along Y-axis as a function of Cassie-Baxter angle ($\Theta_{CB}$) in **Fig. 4**. The initial contact angle of pristine polyimide (110°) is taken as a reference ($\Theta_0$) as shown in **Fig. 4(a)**. **Fig. 4(b)** shows a contact angle of 123° for a laser-patterned surface with N = 1 at 290 nJ energy. The increase in contact angle by 13° is due to the increased surface roughness caused by formation of blisters. Since the film thickness was 1.3 µm, we lowered the energy to avoid rupturing of the film [13]. We decreased the energy to 260 nJ for N = 2 and 220 nJ for N = 4. These surfaces showed contact angles of 128° and 145° (**Fig. 4(c)** and **Fig. 4(d)**). These results indicate that the further increase in hydrophobicity in polymers when irradiated in back-illumination are due to increase in (i) surface roughness and (ii) density of air pockets.

The liquid contact area fraction (f) as given by the equation (above) assumes a homogenous rough surface with square blisters and is a function of both roughness and density of air pockets underneath the water droplet. The calculated liquid contact area fractions for patterned surfaces of 1, 2, and 4-shots are 0.7, 0.6, and 0.3 with a roughness of 225 nm, 125 nm, and 59 nm, respectively. Though the surface roughness for the patterned surface of N = 2 was higher than the patterned surface of N = 1, the density of air pockets for the latter case (N = 1) was slightly lower than the density of air pockets for the patterned surface of N = 1 due to a change in pitch ($\Delta X$). The pitch for N = 1 and N =2 surfaces were 10 µm and 9 µm respectively. The higher the pitch, the lower is the density of air pockets. For the patterned surface of N = 4, the density of air pockets was even higher when the pitch was reduced to half ($\Delta X$ = 5 µm). These quantitative results clearly indicate that the interplay between the surface roughness and density of air-pockets play a key role for the contact angle.

 Based on our contact angle measurements, we propose a two-step LIFT procedure to obtain contaminant-free transfer of sensitive materials such as cells and organelles. In the first step the surface roughness of an existing hydrophobic material is structures in back-illumination to enhance its hydrophobicity. In the second step the nanostructured surface is then irradiated with another pulse for gentle desorption.

## CONCLUSIONS

We studied ultrafast-laser-induced photochemical and morphological changes in polyimide thin films using FIB, XPS, AFM, and water contact angle measurements. Upon nonlinear interaction of ultrafast light with a polyimide film at the interface, the thin film was locally transformed and delaminated from the substrate, leading to formation of a blister and an embedded modified layer with a different phase contrast beneath the film. Both XPS and FIB measurements show that the chemical changes are confined to the glass-polyimide interface. The morphological change due to blister formation is evident through FIB measurements. Blister-patterned surfaces show increased contact angle of water droplets due to increased surface roughness and therefore, increased hydrophobicity. By isolating morphological changes from chemical modification, we show that surface roughness increases hydrophobicity and thereby decreases surface adhesion.

The reduced adhesion that we have observed may be useful for contaminant-free LIFT since laser-induced chemical changes can be confined near the glass-polymer interface while morphological changes can help materials to gently desorb. In addition, confining the region of chemical change is essential for 3D optical data storage based on laser-induced fluorescence from polymers while isolated morphological changes may help create water repellent surfaces for applications in de-icing and defrosting.

## Acknowledgments

All authors acknowledge financial support from Natural Sciences and Engineering Research Council of Canada (NSERC), Ontario Centres of Excellence (OCE), and Fluidigm Canada, Markham, Ontario. Alan T. K. Godfrey acknowledges financial support from NSERC's Postgraduate Scholarship - Doctoral and University of Ottawa's Excellence Scholarship. We acknowledge Dr. Maohui Chen for training in atomic force microscopy, Dr. Choloong Hahn for FIB measurements, Tony Olivieri for training related to polyimide film fabrication, and Dr. Alexander Sander for acquiring XPS spectra. We acknowledge the help received from Charbel Atallah, PhD student, Department of Chemical and Biological Engineering, University of Ottawa for water drop contact angle measurements. We acknowledge help received from lab engineer Yu-Hsuan Wang.

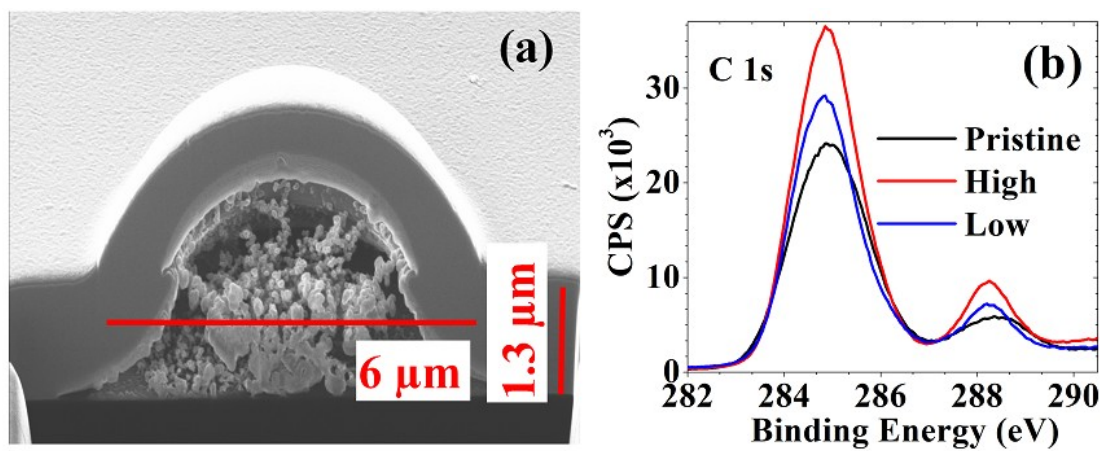

**Fig. 1: (a)** FIB image of a blister fabricated with ~ 130 nJ pulse energy focused by a 0.4 NA objective. The height and diameter of the blister are 1.6 µm and 6 µm respectively. Chemical changes induced by a femtosecond laser pulse (an embedded layer beneath the film) are seen with a different phase contrast. **(b)** XPS of C(1S) envelope for pristine (black) and laser-modified polyimide (high and low irradiation doses shown in red and blue). High and low irradiation dosed regions were patterned using a 0.2 NA objective at 870 nJ and 330 nJ pulse energies, scan speeds of 4.5 mm/s and 3 mm/s, and line spacings of 12 µm and 8 µm respectively at a fixed laser repetition rate of 500 Hz.

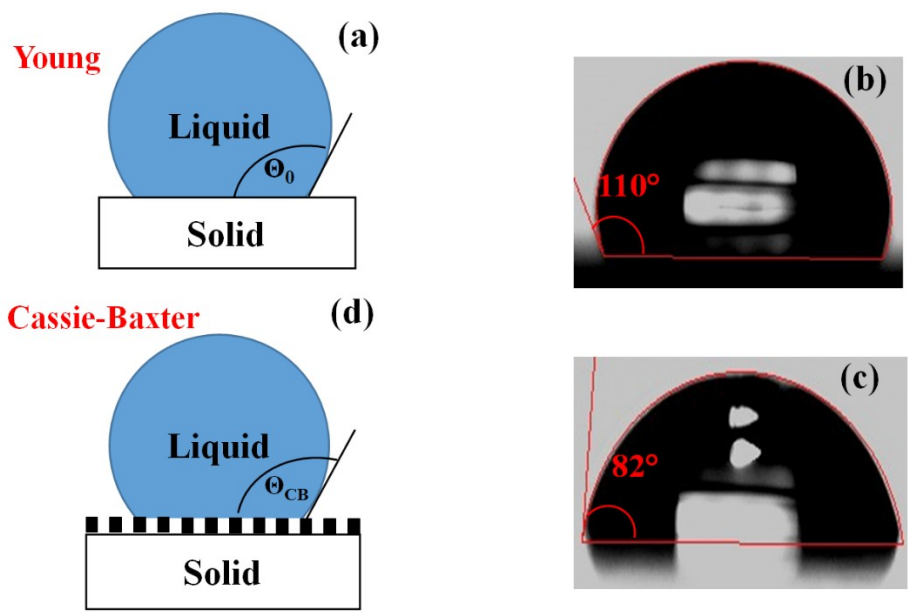

**Fig. 2: (a)** A flat substrate showing Young's contact angle ($\Theta_0$). **(b)** and **(c)** show contact angles of 110° and 82° of pristine polyimide films baked at 300°C and 180°C for 30 mins. **(d)** Young's contact angle ($\Theta_0$) changes to Cassie-Baxter angle ($\Theta_{CB}$) for hydrophobic surfaces upon addition of surface roughness. Additional roughness increases its initial hydrophobicity ($\Theta_{CB} > \Theta_0$).

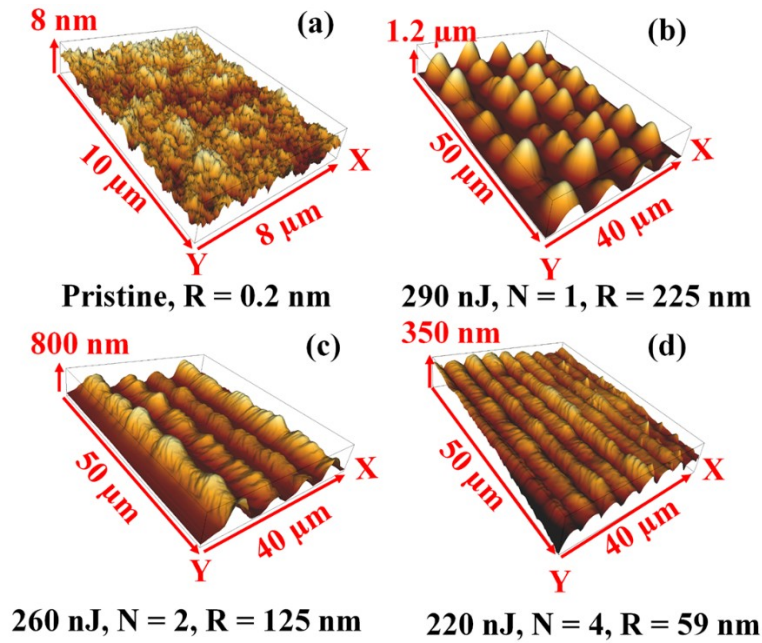

**Fig3: (a)** AFM image of a pristine polyimide surface with average roughness of 0.2 nm with a maximum surface height of 8 nm. Surface topographies shown in **(b), (c),** and **(d)** are the laser-patterned surfaces of polyimide fabricated with different energies (E), no. of shots (N), roughness (R), at line spacings (ΔX) of 10 µm, 9 µm, and 5 µm, respectively.

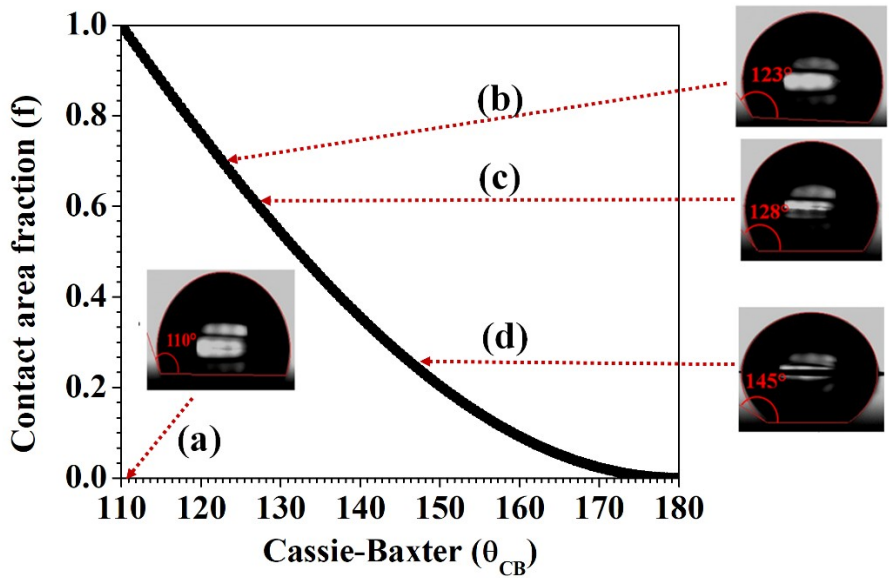

**Fig. 4:** Plot of liquid contact area fraction (f) dependence on the Cassie-Baxter angle ($\Theta_{CB}$). Water contact angle measurement of **(a)** pristine polyimide (110°). Here, both the Cassie-Baxter and Young's contact angle are equal ($\Theta_{CB} = \Theta_0$). Laser-patterned surface with **(b)** N = 1 (123°), **(c)** N = 2 (128°), and **(d)** N = 4 (145°).

# Surface Adhesion of Back-Illuminated Ultrafast Laser-Treated Polymers: Supplemental

Deepak L. N. Kallepalli*, Alan T. K. Godfrey, Jesse Ratté, André Staudte, Chunmei Zhang, and P. B. Corkum*

Joint Attosecond Science Laboratory, University of Ottawa and National Research Council of Canada, 100 Sussex Dr., Ottawa K1N 6N5, Canada

Deepak.Kallepalli@uottawa.ca, pcorkum@uottawa.ca

## I.     Characterization details

The interior modification confined to the interface was studied using focused ion beam (FIB) measurements. Zeiss's ORION NanoFab multi-column (GFIS, and Gallium-FIB) Helium Ion Microscope (HIM) and Focused Ion Beam (Gallium- FIB) were used to study the blistered samples. Gallium ions were used to dissect the blisters and Helium ions were used to image their cross-sections. Prior to performing dissection, the samples were coated with 30 nm of aluminium to protect from excessive damage due to gallium ion exposure. All AFM images were taken using the Nanowizard® II BioAFM (JPK Instruments, Berlin, Germany) mounted on an Olympus IX81 inverted confocal microscope, operating in contact mode. Silicon nitride cantilevers (DNP-S, Veeco, CA) were used in contact mode imaging. XPS measurements were performed using Al Kα as an excitation source with energy of 1,486 eV.

Wettability tests were performed with a goniometer (VCA Optima AST Products Inc.). The instrument had an in-line camera arranged with a syringe for dispensing water. Our measurements were performed with a 1 μL droplet. The laser-patterned surfaces were placed underneath the syringe; the patterned surface was slowly moved up manually and monitored through the camera viewing window of the software until it touched the droplet hanging from the syringe. Once the droplet was placed on the patterned surface, both advancing and receding contact angles were manually adjusted to match the droplet contour viewed through the camera in the software. The difference in between advancing and receding contact angles were small. Hence, the averaged contact angle was shown in images.

## II.     XPS ANALYSIS FOR O(1S) and N(1S) ENVELOPES

**Figure S1(a)** and **Figure S1(b)** show the XPS spectra for O(1S) and N (1S) envelopes for pristine (black) and laser-irradiated polyimide samples (blue and red). The O(1S) and N(1S) envelopes showed peaks at 532 eV and 400 eV (black curves in **Fig. S1(a)** and **Fig. S1(b)**) respectively [1]. The laser-modified polyimide surfaces were corrected for charge compensation and compared with pristine. We did not observe any new bands and peak shifts, indicating that the back-illuminated laser-irradiated polyimide did not undergo any chemical changes on its surface.

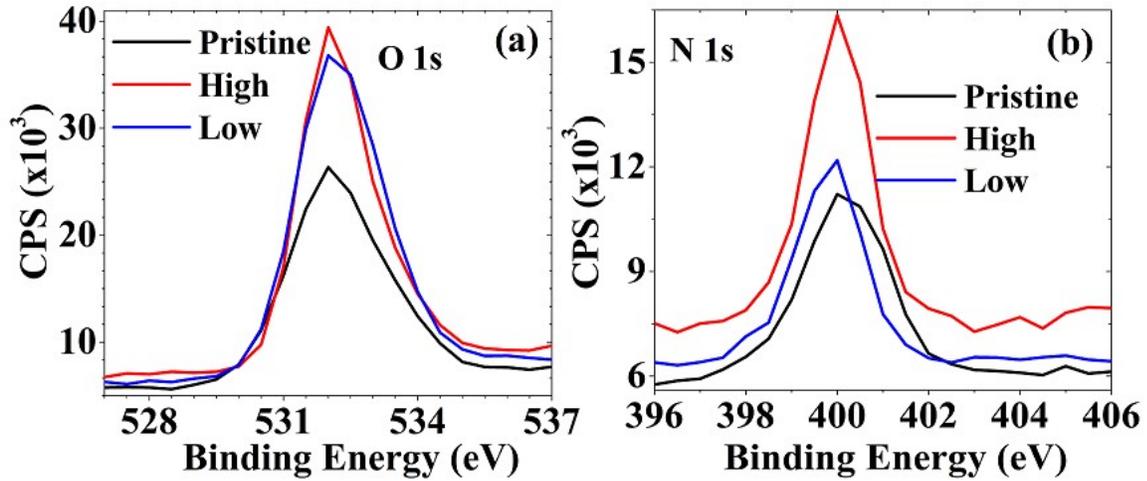

**FigS1: (a)** O(1S) **(b)** N(1S) envelopes for pristine (black) and laser-modified polyimide (high and low irradiation doses shown in red and blue). High and low irradiation dosed regions were patterned using a 0.2 NA objective at 870 nJ and 330 nJ pulse energies, scan speeds of 4.5 mm/s and 3 mm/s, and line spacings of 12 μm and 8 μm respectively at a fixed laser repetition rate of 500 Hz.